\documentclass{jpsj2}
\usepackage{bm}

\def\runauthor{Online-Journal Subcommittee}

\newcommand{\prb}{Phys. Rev. B }
\newcommand{\pra}{Phys. Rev. A }
\newcommand{\prl}{Phys. Rev. Lett. }

\title{Theory of spin qubits in nanostructures}

\author{%
Bj{\"o}rn Trauzettel, Massoud Borhani, Mircea Trif, and Daniel Loss}

\inst{%
Department of Physics and Astronomy, University of Basel,
Klingelbergstrasse 82, 4056 Basel, Switzerland
}

\recdate{August 2007}

\abst{%
We review recent advances on the theory of spin qubits in nanostructures. We
focus on four selected topics. First, we show how to form spin qubits in
the new and promising material graphene. Afterwards, we discuss spin
relaxation and decoherence in quantum dots. In particular, we demonstrate how
charge fluctations in the surrounding environment cause spin decay via
spin--orbit coupling. We then turn to a brief overview of how one can use
electron--dipole spin resonance (EDSR) 
to perform single spin rotations in quantum dots using an
oscillating electric field. The final topic we cover is the spin--spin
coupling via spin--orbit interaction which is an alternative to the
usual spin--spin coupling via the Heisenberg exchange interaction.}

\kword{%
spin qubits, graphene, decoherence, EDSR, spin--spin coupling}

\begin{document}
\maketitle

\section{Introduction}

Spin qubits in quantum dot nanostructures (and the underlying physics) 
is a rapidly evolving
research area. After the original proposal \cite{loss98} based on
the electron spin in few electron quantum
dots, a number of related research branches have emerged. One of
which deals with alternative ways to form spin qubits in solid
state devices. Here, the aim is to find spin qubit realizations
that either couple only weakly to the environment or that are easy to
manipulate. A popular example with a growing interest in the spin
qubit community is two-spin qubits (where the spin qubit
corresponds to singlet and triplet states of two electrons).
\cite{levy02,pett05,tayl05,burk06,hans07} Further examples are
many-spin cluster qubits composed of antiferromagnetically-coupled
spin chains \cite{meie03a,meie03b} and spin qubits in
magnetic molecules. \cite{leue01,lehm07} Another active research
field is devoted to coherence properties of spin qubits. Here,
different aspects of spin relaxation and spin dephasing have been
quantitatively analyzed for different dissipation channels. The
most dominant ones are spin--orbit interaction, coupling the spin
to lattice vibrations \cite{khae00,khae01,golo04,amasha06} and other charge
fluctuations \cite{bor06} as well as the
hyperfine interaction of the electron spin with the surrounding
nuclear spins. \cite{burk99,erli01,khae02,Merkulov02,cois04,klaus06}

The success of spin qubits in nanostructures is substantially due
to the major experimental breakthroughs that have been achieved in
recent years (for recent review articles on spin qubits
see Refs.~\citeonline{cerle05,coish06,hansrev06}). 
After pioneering experiments on few electron quantum
dots \cite{taruc96,ono02,fujis02,hayas03}, 
a first step towards the realization of
quantum computing with the spin of electrons in quantum dots 
has been made in single-shot measurements of the electron spin. 
\cite{Hanson03,elze04,hans05} Subsequently, a
coherent two-qubit gate (the $\sqrt{{\rm SWAP}}$ gate) has been
realized. \cite{pett05} Recently, coherent single spin rotations
have been demonstrated via electron spin resonance techniques
using pulsed magnetic fields. \cite{kopp06} Thus, all single- and
two-qubit operations required for universal quantum computing have
been realized in spin qubits based on the original idea.
\cite{loss98} However, the time scales needed to operate 
single-qubit gates in spin qubits hosted in lateral quantum dots
in GaAs/AlGaAs heterostructures \cite{elze04,hans05,kopp06}
are still quite long as compared to the decoherence time $T_2 \sim 1-10\mu$s. 
(Note that
this is not the case for two-qubit operations which can be performed as
fast as 180 ps for the $\sqrt{{\rm SWAP}}$ gate \cite{pett05}.) 
The ratio of the operation time and the decoherence time should be
of the order of $10^4$ to be able to do fault-tolerant quantum
computing. The decoherence time is currently limited by the
hyperfine interaction of the electron spin with the surrounding
nuclear spins \cite{ono04,johns05,koppe05}. 
Therefore, it is desirable to form spin qubits in
other materials where spin relaxation and spin decoherence are
less efficient than in GaAs/AlGaAs heterostructures. Two examples (which have
already been realized) are few-electron quantum dots
in carbon nanotubes \cite{mason04,bierc05,sapma06,graeb06} 
as well as semiconductor nanowires \cite{fasth05,pfund07}. We discuss
below another interesting example, namely spin qubits in graphene, where
spin relaxation and spin decoherence mechanisms are expected to be
weaker than in GaAs-based devices (for recent review articles on graphene see
Refs.~\citeonline{castro06,geim07,katsn07}).

The article is organized as follows: In Sec.~\ref{sec_2},
we explain in detail our recent proposal to form spin qubits in
graphene quantum dots. Many of the aspects discussed in that
section equally apply to spin qubits based on carbon nanotube
quantum dots. In Sec.~\ref{sec_3}, different aspects of spin
relaxation and decoherence are reviewed. Afterwards, in
Sec.~\ref{sec_4}, recent ideas to use EDSR to form single spin
rotations in quantum dots are discussed. In Sec.~\ref{sec_5}, we
show how spin--spin coupling can be achieved via spin--orbit
interaction. Finally, we conclude in Sec.~\ref{sec_6}.

\section{Spin qubits in graphene quantum dots}
\label{sec_2}

It is generally believed that carbon-based materials such as
nanotubes or graphene are excellent candidates to form spin qubits
in quantum dots. This is because spin-orbit coupling is weak in
carbon (due to its relatively low atomic weight) \cite{ando00,min06,huert06}, 
and
because natural carbon consists predominantly of the zero-spin
isotope $^{12}$C, for which the hyperfine interaction is absent.
In this section, we review how to form spin qubits in graphene.
\cite{trau07} A crucial requirement to achieve this goal is to
find quantum dot states where the usual valley degeneracy is
lifted. We show that this problem can be avoided in quantum dots
with so-called armchair boundaries.
We furthermore show that spin qubits in graphene can not only be
coupled (via Heisenberg exchange) between nearest neighbor quantum
dots but also over  long distances. This remarkable feature is a
direct consequence of the Klein paradox being a distinct property
of the quasi-relativistic spectrum of graphene. \cite{kats06}

Two fundamental problems need to be overcome before graphene can
be used to form spin qubits and to operate one or two of them in
the standard way. \cite{loss98,burk99} (i) It is difficult to
create a tunable quantum dot in graphene because of the absence of
a gap in the spectrum. \cite{chei06,kats06}. (ii) Due to the
valley degeneracy that exists in graphene \cite{seme84,divi84}, it
is non-trivial to form two-qubit gates using Heisenberg exchange
coupling for spins in neighboring dots.
Several attempts have
been made to solve the problem (i) \cite{nils06,silv07,dema07,milto07,staub07} 
(without having problem (ii) in mind).
We have recently proposed a setup which solves both problems (i)
and (ii) at once. \cite{trau07} In particular, we assume {\it
semiconducting armchair} boundary conditions to exist on two
opposite edges of the sample. It is known that in such a device
the valley degeneracy is lifted \cite{brey06,twor06}, which is the
essential prerequisite for the appearance of Heisenberg exchange
coupling for spins in tunnel-coupled quantum dots, and thus for
the use of graphene dots for spin qubits.

\begin{figure}[t]
\begin{center}
\includegraphics[width=6cm]{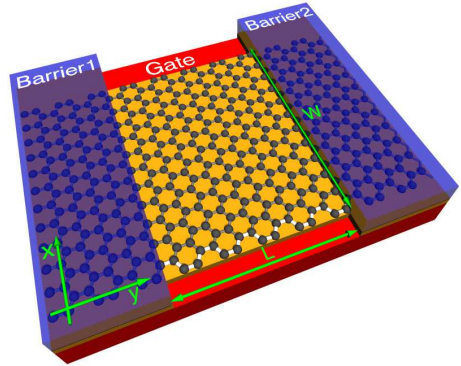}
\end{center}
\caption{(Color online) A ribbon of graphene with semi-conducting armchair
boundaries is schematically shown. Two barrier gates (blue) define
the rectangular size of the quantum dot (with width $W$ and length
$L$). A back gate (red) allows one to shift the energy levels in
the dot.} \label{setup}
\end{figure}

We now discuss bound-state solutions in the appropriate setup,
which are required for a localized qubit.
We first concentrate on a single quantum dot which is assumed to
be rectangular with width $W$ and length $L$, see
Fig.~\ref{setup}.
The basic idea of forming the dot is to take a ribbon of graphene
with semiconducting armchair boundary conditions in $x$-direction
and to electrically confine particles in $y$-direction.

The low energy properties of electrons (with energy $\varepsilon$
with respect to the Dirac point) in such a setup are described by
the 4x4 Dirac equation
\begin{equation} \label{Dirac}
\frac{\hbar v}{i} \begin{pmatrix}
\sigma_{x}\partial_{x}+\sigma_{y}\partial_{y}&0\\
0&-\sigma_{x}\partial_{x}+\sigma_{y}\partial_{y}
\end{pmatrix}\Psi+\mu(y)\Psi =\varepsilon\Psi,
\end{equation}
where the electric gate potential is assumed to vary stepwise,
$\mu(y) = \mu_{\rm gate}$ in the dot region (where $0\le y\le L$),
and $\mu(y) = \mu_{\rm barrier}$ in the barrier region (where
$y<0$ or $y>L$). In Eq.~(\ref{Dirac}), $\sigma_x$ and $\sigma_y$
are Pauli matrices (denoting the sublattices in graphene). 
The four component spinor envelope wave
function $\Psi =
(\Psi^{(K)}_A,\Psi^{(K)}_B,-\Psi^{(K')}_A,-\Psi^{(K')}_B)$ varies
on scales large compared to the lattice spacing. Here, $A$ and $B$
refer to the two sublattices in the two-dimensional honeycomb
lattice of carbon atoms, whereas $K$ and $K'$ refer to the vectors
in reciprocal space corresponding to the two valleys in the
bandstructure of graphene.
The appropriate semiconducting armchair boundary conditions for
such a wave function can be written as ($\alpha = A,B$)
\cite{brey06}
\begin{eqnarray} \label{bc}
\Psi_\alpha^{(K)}|_{x=0} &=& \Psi_\alpha^{(K')}|_{x=0} , \nonumber
\\
\Psi_\alpha^{(K)}|_{x=W} &=& e^{\pm 2\pi/3}
\Psi_\alpha^{(K')}|_{x=W} .
\end{eqnarray}
These boundary conditions couple the two valleys and are, thus,
the reason why the valley degeneracy is lifted. \cite{foot1} It is
well known that the boundary condition (\ref{bc}) yields the
following quantization conditions for the wave vector $k_x \equiv
q_n$ in $x$-direction \cite{brey06,twor06}
\begin{equation} \label{qn1}
q_n = (n \pm 1/3)\pi/W , \;\; n \in \mathbb{Z} .
\end{equation}
The level spacing of the modes (\ref{qn1}) can be estimated as
$\Delta \varepsilon \approx \hbar v \pi/3W$, which gives $\Delta
\varepsilon \sim  30 \,{\rm meV}$, where we used that $v
\sim 10^6 \, {\rm
  m/s}$ and assumed a quantum dot width of about $W \sim 30 \, {\rm nm}$.
Note that Eq.~(\ref{qn1}) also determines the energy gap for
excitations as $E_{\rm gap} = 2 \hbar v q_0$. Therefore, this gap
is of the order of 60 meV, which is unusually small for
semiconductors. This is a unique feature of graphene that will
allow for long-distance coupling of spin qubits as will be
discussed below.

\begin{figure}[t]
\vspace*{0.5cm}
\begin{center}
\includegraphics[width=6.5cm]{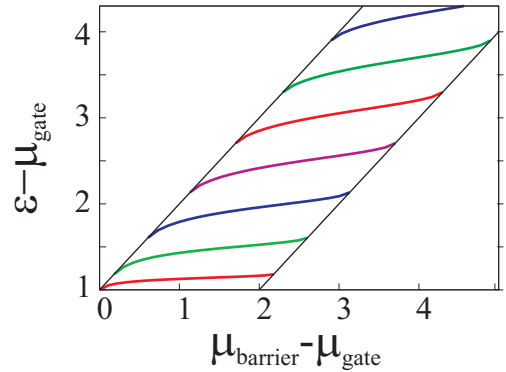}
\caption{\label{fig2} (Color online) Bound-state solutions of a dot with aspect
ratio $q_0 L=\pi L/3W=5$. The diagonal lines indicate the region
in which bound-state solutions do exist given by
Eq.~(\ref{window1}). All energies are taken in units of $\hbar v
q_0$.}
\end{center}
\end{figure}

We now present in more detail the ground-state solutions, i.e.
$n=0$ in Eq.~(\ref{qn1}). The corresponding ground-state energy
$\varepsilon$ can be expressed relative to the potential barrier
$\mu=\mu_{\rm barrier}$ in the regions $y<0$ and $y>L$ as
$\varepsilon=\mu_{\rm barrier} \pm \hbar v
(q_{0}^{2}+k^{2})^{1/2}$.
Here, the $\pm$ sign refers to a conduction band ($+$) and a
valence band ($-$) solution to Eq.~(\ref{Dirac}).
For bound states to exist and to decay at $y \rightarrow \pm
\infty$, we require that $\hbar v q_0 > |\varepsilon -\mu_{\rm
barrier}|$, which implies that the wave vector $k_y \equiv k$ in
$y$-direction, given by
\begin{equation} \label{kbarrier}
k = i \sqrt{q_0^2 - ((\varepsilon -\mu_{\rm barrier})/\hbar v)^2}
,
\end{equation}
is purely imaginary.
In the dot region ($0 \leq y \leq L$), the wave vector $k$ in
$y$-direction is replaced by $\tilde{k}$, satisfying
$\varepsilon=\mu_{\mathrm{gate}}\pm \hbar v
(q_0^{2}+\tilde{k}^{2})^{1/2}.$ Again the $\pm$ sign refers to
conduction and valence band solutions. (In the following, we focus
on conduction band solutions to the problem.)
In the energy window
\begin{equation} \label{window1}
|\varepsilon-\mu_{\rm gate}| \ge \hbar v q_0 >
|\varepsilon-\mu_{\rm barrier}| ,
\end{equation}
the bound state energies are given by the solutions of the
transcendental equation
\begin{equation} \label{trans1}
\tan(\tilde{k} L) = \frac{\hbar v \tilde{k} \sqrt{(\hbar v q_0)^2
- (\varepsilon -\mu_{\rm barrier})^2}}{(\varepsilon - \mu_{\rm
  barrier})(\varepsilon - \mu_{\rm gate}) - (\hbar v q_0)^2} .
\end{equation}
We show a set of solutions to Eq.~(\ref{trans1}) for a dot with
aspect ratio $q_0 L=\pi L/3W=5$ in Fig.~\ref{fig2}.

We now turn to the case of two coupled graphene quantum dots,
separated by a potential barrier, each dot filled with a single
electron.
It is interesting to ask whether the spins ${\bf S}_i$ of these
two electrons ($i=1,2$) are coupled through an exchange coupling,
$H_{\rm exch}=J{\bf S}_1\cdot{\bf S}_2$, in the same way as for
regular semiconductor quantum dots \cite{burk99}, because this
coupling is, in combination with single-spin rotations, sufficient
to generate all quantum gates required for universal quantum
computation \cite{loss98}.
The exchange coupling is based on the Pauli exclusion principle
which allows for electron hopping between the dots in the spin
singlet state (with opposite spins) of two electrons, but not in a
spin triplet (with parallel spins), thus leading to a
singlet-triplet splitting (exchange energy) $J$.

However, a singlet-triplet splitting $J\neq 0$ only occurs if the
triplet state with two electrons on the same dot in the ground
state is forbidden, i.e., in the case of a single
\textit{non-degenerate} orbital level.
This is a non-trivial requirement in a graphene structure, as in
bulk graphene, there is a two-fold orbital degeneracy of states
around the points $K$ and $K'$ in the first Brillouin zone.
This valley degeneracy is lifted in our case of a ribbon with {\it
  semiconducting armchair
edges}, and the ground-state solutions determined by
Eq.~(\ref{trans1}) are in fact non-degenerate \cite{foot2}.
The magnitude of the exchange coupling within a Hund-Mulliken
model is \cite{burk99} $J= (-U_H+ (U_H^2+16t_H^2)^{1/2})/2+V$,
where $t$ is the tunneling (hopping) matrix element between the
left and right dot, $U$ is the on-site Coulomb energy, and $V$ is
the direct exchange from the long-range (inter-dot) Coulomb
interaction. The symbols $t_H$ and $U_H$ indicate that these
quantities are renormalized from the bare values $t$ and $U$ by
the inter-dot Coulomb interaction.

For $t \ll U$ and neglecting the long-ranged Coulomb part, this
simplifies to the Hubbard model result $J=4t^2/U$ where $t$ is the
tunneling (hopping) matrix element between the left and right dot
and $U$ is the on-site Coulomb energy.
In the regime of weak tunneling, we can estimate $t \approx
\varepsilon  \int \Psi_L^\dagger(x,y)\Psi_R(x,y)  dx \, dy$, where
$\Psi_{L,R}(x,y)=\Psi(x,y \pm (d+L)/2)$ are the ground-state
spinor wave functions of the left and right dots and $\varepsilon$
is the single-particle ground state energy.
Note that the overlap integral vanishes if the states on the left
and right dot belong to different transverse quantum numbers
$q_{n_L}\neq q_{n_R}$.

For the ground state mode, we have $n_L=n_R=0$, and the hopping
matrix element can be estimated for $d\gtrsim L$ as
\begin{equation} \label{t}
t \approx 4 \varepsilon  \alpha_0 \delta^*_0 W d z_{0,k} \exp(-d
|k|) ,
\end{equation}
where $\alpha_0$ and $\delta_0$ are wave function amplitudes (with
dimension 1/length), see Ref.~\citeonline{trau07} for more details. As
expected, the exchange coupling decreases exponentially with the
barrier thickness, the exponent given by the ``forbidden''
momentum $k$ in the barrier, defined in Eq.~(\ref{kbarrier}).

The values of $t$, $U$, and $J$ can be estimated as follows. The
tunneling matrix element $t$ is a fraction of $\varepsilon \sim
30\,{\rm meV}$ (for a width of $W \sim 30 \,{\rm nm}$), we
obtain that $t \sim 0.5 \dots 2.5 \,{\rm meV}$. The value for
$U$ depends on screening which we can assume to be relatively weak
in graphene \cite{divi84}, thus, we estimate, e.g., $U \sim 10
\,{\rm meV}$, and obtain $J \sim 0.1 \dots 1.5 \,{\rm meV}$. (Note that
this rough estimate would correspond to very fast switching times $\tau_s \sim
\hbar/J \sim 1 \dots 10$ps for the $\sqrt{{\rm SWAP}}$ operation.) 

\begin{figure}[t]
\begin{center}
\includegraphics[width=12cm]{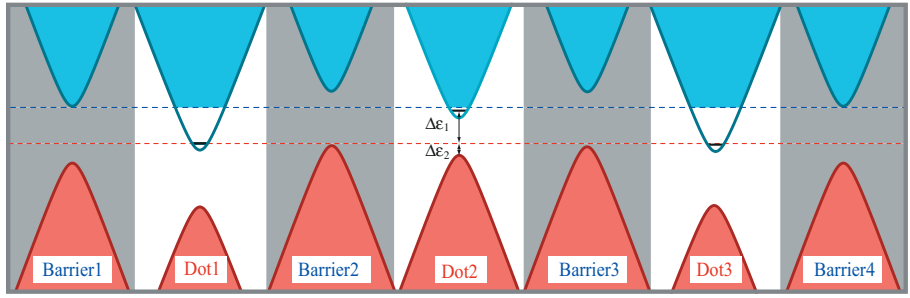}
\vspace*{0.5cm} \caption{\label{fig3} (Color online) 
The energy bands of a triple
quantum dot setup are shown in which dot 1 and dot 3 are strongly
coupled via cotunneling processes through the valence bands of
barrier 2, barrier 3, and dot 2. The center dot 2 is decoupled by
detuning. The energy levels are chosen such that $\Delta
\varepsilon_2 \ll \Delta \varepsilon_1$. The triple dot example
illustrates that in a line of quantum dots, it is possible to
strongly couple any two of them and decouple the others by
detuning. This is a unique feature of graphene and cannot be
achieved in semiconductors such as GaAs that have a much larger
gap (after Ref.\cite{trau07}) .}
\end{center}
\end{figure}

For the situation with more than two dots in a line, it turns out
that we can couple any two of them with the others being decoupled
by detuning. In Fig.~\ref{fig3}, we illustrate the situation of
three dots in a line where the left and the right dot are strongly
coupled and the center dot is decoupled by detuning. The tunnel
coupling of dot 1 and dot 3 is then achieved via Klein tunneling
through the valence band of the two central barriers and the
valence band of the center dot. It is important for the
long-distance coupling that the exchange coupling of qubit 1 and
qubit 3 is primarily achieved via the valence band and not via the
qubit level of the center dot -- leaving the qubit state of dot 2
unchanged. Using the standard transition matrix approach, we can
compare the transition rate of coupling dot 1 and dot 3 via the
continuum of states in the valence band of the center dot (which
we call $\Gamma_{\rm VB}$) with the transition rate via the
detuned qubit level of the center dot (which we call $\Gamma_{\rm
QB}$). We obtain for the ratio \cite{trau07}
\begin{equation} \label{ratio}
\Gamma_{\rm VB}/\Gamma_{\rm QB} \approx (L/W) \ln (4 \Delta/E_{\rm
gap}) ,
\end{equation}
where $\Delta \sim 6 \,{\rm eV}$ is the band width of graphene.
Therefore, by increasing the aspect ratio $L/W$, it is possible to
increase the rate $\Gamma_{\rm VB}$ with respect to $\Gamma_{\rm
QB}$. For $L/W=2$ and $E_{\rm gap} \sim 60 \,{\rm meV}$, we find that
$\Gamma_{\rm VB}/\Gamma_{\rm QB} \sim 12$, meaning that the
qubit level in dot 2 is barely used to couple dot 1 and dot 3.
This is a unique feature of graphene quantum dots due to the small
and highly symmetric band gap.

%

\section{Spin relaxation and decoherence in quantum dots}

\label{sec_3}
Phase coherence of spins in quantum dots (QDs) is of central importance for
spin-based quantum computation 
in the solid state, however, the mechanisms of spin decoherence for extended
and localized electrons are rather different. 
Different mechanisms of spin relaxation in QDs have been considered, such as
spin-phonon coupling via spin-orbit (SO) interaction
\cite{khae00,khae01,golo04,bor06} or hyperfine interaction  
\cite{erli01}, and direct hyperfine coupling
\cite{burk99,khae02,Merkulov02,cois04,klaus06}. 
In this section, we consider the spin relaxation and decoherence in quantum
dots 
due to the coupling to phonons and charge fluctuations in the surrounding
environment \cite{golo04,bor06}. We show how SO interaction
couples the electron spin to these types of fluctuations by deriving an effective Hamiltonian
for the spin subspace. Throughout this section, we consider only the leading
contribution to SO interaction in two dimensional systems 
(commonly called linear-in-$p$ SO interaction)
\begin{equation}
H_{SO} = \beta(-p_x\sigma_x + p_y\sigma_y) + \alpha(p_x\sigma_y - p_y\sigma_x),\label{RDSOI}
\end{equation} 
where $\alpha$ and $\beta$ are Rashba and Dresselhaus coefficients,
respectively. Both coefficients have been measured recently in GaAs/InGaAs
quantum wells using optical detection schemes. \cite{meier07} 
(Note that in this and in the following sections the Pauli
matrices denote the electron spin whereas in the previous section it was the
sublattice index of graphene.) Moreover, we assume that the temperature is
the smallest energy scale in the system and the Zeeman energy is less than the
orbital quantization in the QD, $k_B T \ll E_Z \ll \hbar\omega_0$. 

\begin{figure}
\begin{center}
\includegraphics[angle=0,width=0.5\textwidth]{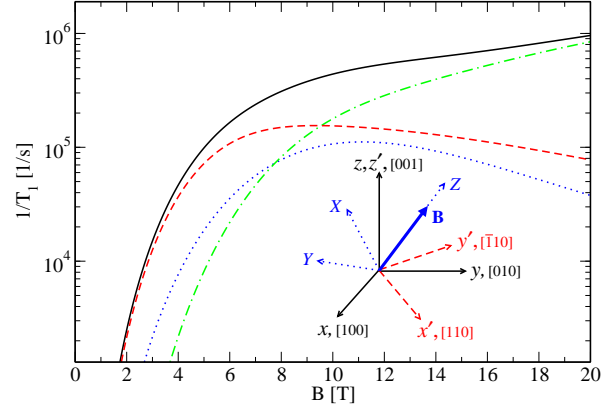}
\caption{\small (Color online)
Solid curve: The relaxation rate $1/T_1$ due to phonons as a function of an in-plane ${\bm B}$ for a GaAs
QD with $\hbar \omega_0 = 1.1$ meV, $\lambda_{so}=\hbar/m^*\beta=1\;\mu$m, and $\alpha=0$. Dashed (dotted)
curve: Contribution of the piezoelectric mechanism with transverse (longitudinal) phonons. Dot-dashed curve:
Contribution of the deformation potential mechanism. Different coordinate frames are used for the relaxation
rate calculations; $(x,y,z)$ are the main crystallographic axes and $(x',y',z')$ are defined as 
$x'=(x+y)/\sqrt{2}$, $y'=(y-x)/\sqrt{2}$ and $z'=z$.
}
\label{vitaly}
\end{center}
\end{figure}

{\it {Phonon contribution}} - Lattice vibrations perturb the confining potential $U({\bm r})$ of the dot and
these fluctuations couple to the electron spin in the QD via the spin--orbit interaction. At low temperatures,
the effective Hamiltonian for the electron spin is given by \cite {golo04}
\begin{eqnarray}
H_{\rm eff} &=& \frac{1}{2} g \mu_B \left[{{\bm B}} + \delta
{{\bm B}}(t) \right ]
\cdot {{\bm \sigma}},\label{Heff}\\
\delta {{\bm B}}(t) &=&
2{\bm B}  \times  {\bm \Omega}(t),\label{Omega}
\end{eqnarray} 
where ${\bm B}$ is the applied magnetic field and ${\bm \Omega}(t)$ is the quantum fluctuating field due
to the coupling to phonons. Eqs. ({\ref {Heff},\ref{Omega}}) show an important
result:
In first order in SO interaction, there can be only transverse fluctuations
of the effective magnetic field, i.e., $\delta {{\bm B}}(t) \cdot {\bm B} =0
$, and the coupling is proportional to the $B$-field itself.
The former property holds true for spin coupling to any fluctuations, be it the noise of a gate voltage or
coupling to particle-hole excitations in a Fermi sea. Consequently, there is no pure dephasing and
$T_2 = 2 T_1$ for arbitrarily large Zeeman splitting, in contrast to the naively expected case
$T_2 \ll T_1$, where $T_1$ is the longitudinal relaxation time (or simply relaxation time) 
and $T_2$ is the transverse relaxation time (decoherence time) of the spin.
After averaging over the phonon bath, we find that the spin decay rate has a non-trivial magnetic 
field dependence; we do not present here the full analytic expression for the
magnetic field  dependence of the spin decay rate and refer to the 
original work instead \cite{golo04}. We plot $1/T_1$ as a function of in-plane 
 ${\bm B}$ for $\alpha = 0$ (only Dresselhaus spin--orbit interaction), see Fig.(\ref{vitaly}).
In agreement with experiment \cite{Hanson03}, $1/T_1$ shows a plateau in a wide range of $B$ fields,
due to a crossover from piezoelectric-transverse (dashed curve) to the deformation potential 
(dot-dashed curve) mechanism of electron-phonon interaction. Note that if $\alpha = \beta$ and 
${\bm B} \parallel y'$ then $1/T_1$ vanishes 
(the same is true for $\alpha = -\beta$ and ${\bm B} \parallel x'$), where
$x'\equiv[110]$ and $y'\equiv[\bar110]$. A detailed measurement of the
$B$-field dependence was reported recently \cite{amasha06} giving very good
agreement with theory \cite{golo04}. Quite remarkably, the largest measured
$T_1$ times exceed 1 second \cite{amasha06}. 

\begin{figure}
\begin{center}
\includegraphics[angle=0,width=0.45\textwidth]{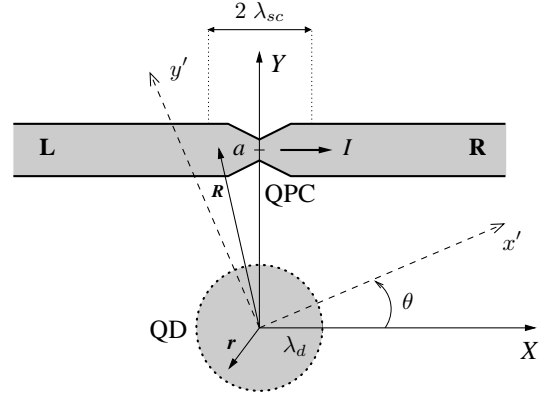}
\caption{\small
Schematic of the quantum dot (QD) coupled to a QPC.
The $(X,Y)$ frame gives  the setup orientation, left (L) and right (R) leads,
with respect to the crystallographic  directions
$x'\equiv[110]$ and $y'\equiv[\bar110]$.
The dot has a radius $\lambda_d$ and is located at a distance $a$ from
the QPC. The vector ${\bm R}$ describes the QPC electrons and ${\bm r}$
refers to the  coordinate of the electron in the dot.
The noise of the QPC current $I$ perturbs the electron spin on the dot
via the spin--orbit interaction.
}
\label{QPC}
\end{center}
\end{figure}

{\it {Quantum point contact (QPC) contribution}} - Charge fluctuations in the surrounding environment of the QD cause spin decay. Here,
we consider one of these sources, a nearby functioning QPC (see Fig.\ref{QPC}), in which
the charge couples to the spin via spin--orbit interaction in the presence of a magnetic field.
The effective Hamiltonian for the electron spin looks the same as Eq.(\ref{Heff}), but here, the
origin of ${\bm \Omega}(t)$ is the electron shot noise in the QPC and its functional dependence on the 
system parameters is different from the phonon case.
There are two mechanisms which contribute to the spin relaxation rate $1/T_1$: The electron-hole excitations
in the QPC Fermi leads and the electron shot noise in the QPC \cite{bor06}. 
In the regime with high bias voltages 
$\Delta \mu$ applied to the QPC, the latter is the dominant one. 
To go further, we assume that the applied magnetic
field  ${\bm B}$ is in-plane and along $x'$ and we obtain
$( E_Z, T \ll  | \Delta\mu \pm E_Z | \ll \hbar \omega_0)$ \cite{bor06}
\begin{eqnarray}
\frac{1}{T_1}  &\approx&  \frac{8 \pi^2  e^2\hbar^4} {{m^*}^2\kappa^2}
\frac{\nu^2 \lambda_{sc}^4}{ a^6 \lambda_+^2}\frac{E_Z^2 \cos^2\theta}
{(\hbar^2\omega_0^2-E_Z^2)^2} S_{LL},\label{final}\\
S_{LL}&=& \frac{e^2 \Delta \mu}{\pi \hbar}{{\cal T}(1-{\cal T})}.
\end{eqnarray}
Here $\nu =1/2\pi\hbar v_F$ is the density of states per spin and mode in the QPC leads,
$m^*$ is the electron effective mass, $\kappa$ is the dielectric constant, $a$ is the distance
form the QD center to the QPC, $\theta$ is the orientation angle of the QPC on the substrate
 (see Fig.\ref{QPC}), $\lambda_{sc}$ is the Coulomb screening length, 
$\lambda_{\pm}=\hbar/m^*(\beta \pm \alpha)$ are spin-orbit lengths, $\hbar \omega_0$ is the 
orbital quantization energy in the QD, ${\cal T}$ is the transmission coefficient of the QPC,
and $S_{LL}$ is the current shot noise. 
Therefore, in this regime, spin decay rate is linear in bias voltage $\Delta \mu$ and scales as $a^{-6}$.
Moreover, $T_1$ strongly depends on the QPC orientation on the 
substrate (the angle $\theta$ between the axes $x'$ and $X$, see
Fig.~(\ref{QPC})), e.g. 
the non-equilibrium part of the relaxation rate vanishes at 
$\theta = \pi/2$, for an in-plane magnetic field ${\bm B}$ along $x'$.
We conclude that the spin decay rate can be minimized
by tuning certain geometrical parameters of the setup. 
Our results should  also be useful for designing experimental setups such
that the spin decoherence can be made negligibly small 
while charge detection with the QPC is still efficient.

\section{EDSR in quantum dots}
\label{sec_4}

Spin--orbit interaction, although it is one of the main sources of the spin decay in QDs, can
be employed to manipulate the electron spin. Here we show how by using an ac
electric field together with a static magnetic field, one can coherently
rotate the spin of the electron around the Bloch sphere \cite{golo06}. 
The physical mechanism responsible for the spin rotation is the so-called
EDSR \cite{rashb91} 
which is similar to usual Electron Spin Resonance (ESR) \cite{ather73,engel01}
but, in the former case, the oscillating electric field replaces the
oscillating magnetic field in the latter case. 
The main advantage of EDSR to ESR is its experimental convenience.

\begin{figure}
\begin{center}
\includegraphics[angle=0,width=0.45\textwidth]{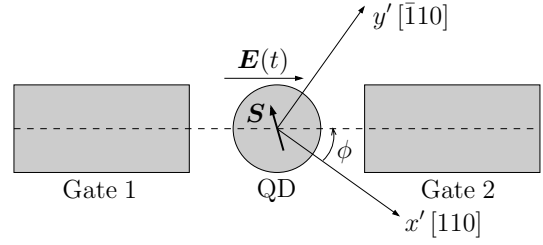}
\caption{\small
Setup for electric field control of spin via the spin--orbit interaction.
The quantum dot (QD) contains a single electron with spin
$\mbox{\boldmath $S$}=(\hbar/2)\mbox{\boldmath $\sigma$}$, deep in the Coulomb
blockade valley, and in the presence of an external static magnetic field
giving rise to a Zeeman splitting $E_Z$. 
The gates $1$ and $2$ are used to generate an alternating
electric field $\mbox{\boldmath $E$}(t)$, which acts via the 
spin--orbit interaction on the electron spin.
As a result, an electric dipole spin resonance (EDSR)
occurs if the frequency of $\mbox{\boldmath $E$}(t)$
is tuned to match the Larmor frequency $\omega_Z=E_Z/\hbar$.
}
\label{EDSRfigure}
\end{center}
\end{figure}

We derive an effective Hamiltonian for the electron spin in the presence of a coherent 
driving {\it ac} electric field $V(\mbox{\boldmath $r$},t)=
e\int^{\bm r} d\mbox{\boldmath $r$}'\cdot
\mbox{\boldmath $E$}(\mbox{\boldmath $r$}',t) \approx  e\mbox{\boldmath $E$}(t)\cdot\mbox{\boldmath $r$}$
(see Fig. \ref{EDSRfigure}).
 Hereby, we use the dipole approximation, ignoring the coordinate dependence of the 
electric field due to the smallness of the dot size compared to the electric field wavelength. 
This leads us to the following effective spin Hamiltonian \cite{golo06}
\begin{eqnarray}
\label{Heffexpl}
H_{\rm eff}&=&\frac{1}{2}g\mu_B{{\bm B}}\cdot\mbox{\boldmath $\sigma$}+
{1\over 2}{\bm h}(t)\cdot\mbox{\boldmath $\sigma$},\\
{\bm h}(t)&=&2g\mu_B{\bm B}\times\mbox{\boldmath $\Omega$}(t),\label{hoftdef}\\
\mbox{\boldmath $\Omega$}(t)&=&\frac{-e}{m^*\omega_0^2}\left(\lambda_-^{-1}E_{y'}(t),
\lambda_+^{-1}E_{x'}(t),0\right),
\label{Omegaexpl}
\end{eqnarray}
where ${\bm E}(t)={\bm E}_0 \sin{(\omega_{ac}t)}$ of amplitude ${\bm E}_0 =E_o(\cos\phi,\sin\phi,0)$ 
and $\phi$ is the angle of  ${\bm E}_0$ with respect to the axis $x'$ (see Fig.\ref{EDSRfigure}). 
Note that the resonance happens when $\omega_{ac}=\omega_Z=E_Z/\hbar$, 
i.e. when the frequency of the driving field
matches the Larmor frequency.
The above Hamiltonian has a similar form to the ESR Hamiltonian, 
except for the fact that the
oscillating electric field plays 
 the role of the {\it ac} magnetic field. Consequently, we can rotate the electron spin around the Bloch 
sphere and build a universal single qubit gate. However, to quantify the efficiency of our EDSR scheme,
 we need to estimate the amplitude of the EDSR 
field, ${\bm h}(t)$, which is proportional to the Rabi frequency $\omega_R$. For GaAs QDs, we assume that
$\lambda_+\approx\lambda_-\approx\lambda_{SO}=8\,\mu{\rm m}$, $|g|=0.44$, $\hbar\omega_0=1\,{\rm meV}$, 
and $E_0=10^2\,{\rm V}/{\rm cm}$, which yields $|{\bm \Omega}|\sim 10^{-3}$. Together with the applied
 magnetic field $B=10\,{\rm T}$, we obtain $\omega_R\sim 10^{8}\,{\rm s}^{-1}$.
We conclude that, with the present QD setups, EDSR enables one to manipulate the electron spin on a time 
scale of $10$ ns, which is considerably shorter than typical spin dephasing
times $T_2 \sim 1 - 10 \mu$s in gated GaAs QDs.  This
mechanism has recently been employed to experimentally rotate the electron
spin in quantum dots \cite{nowack07} with $\pi/2$ rotations as fast as $\sim
55$ns. In a similar experimental setup, hyperfine-mediated gate-driven
electron spin resonance has been observed. \cite{laird07}

Up to now, we have only considered the linear-in-$p$ spin--orbit interaction. However, if the 
two dimensional electron gas (2DEG) has a finite width $d$,
then the so-called $p^3$ terms of the Dresselhaus spin--orbit interaction also come into the play:
\begin{equation}
H_{SO}=\frac{\gamma}{2}\left(p_yp_xp_y\sigma_x-p_xp_yp_x\sigma_y\right),
\label{HSOp3}
\end{equation}
where $\gamma=\alpha_c/\sqrt{2{m^*}^3E_g}$ is the spin-orbit coupling constant,
with $\alpha_c$ ($\sim 0.07$ for GaAs), and $E_g$ the band gap.
Quite remarkably, if the quantum dot potential is harmonic, then the spin does not couple
 to $\mbox{\boldmath $E$}(t)$ in the first order of $H_{SO}$ and zeroth order of $E_Z$ \cite{golo06}.
Thus, for a harmonic confining potential, one is left with the same dominant mechanism as
considered above for the "linear in $p$" terms.
To estimate the strength of the resulting EDSR, we expand in terms of the Zeeman interaction and
note that $\gamma\sim\beta d^2/\hbar^2$, and therefore the amplitude of 
$\mbox{\boldmath $h$}(t)=2g\mu_B{\bm B}\times\mbox{\boldmath $\Omega$}(t)$
is by a factor $d^2/\lambda_d^2\ll 1$ smaller as compared to the corresponding
amplitude of the linear-in-$p$ contributions.

Next we consider a quantum dot with {\it anharmonic} potential $U({\bm r})$ and show that
the $p^3$-terms in Eq.~(\ref{HSOp3}) give rise to a spin-electric coupling
proportional to the cyclotron frequency $\omega_c=eB_z/m^*c$. \cite{golo06}
Since $\hbar\omega_c$ differs parametrically from $E_Z$
($E_Z/\hbar\omega_c=gm^*B/2mB_z$), the $p^3$-terms can be as
significant as the $p$-terms, provided $E_Z/\hbar\omega_c\lesssim d^2/\lambda^2$, which is
realistic for GaAs quantum dots. As an example, we consider $U( r)=m^*\omega_0^2r^2/2\;+\;\eta r^4$ 
, where $\eta$ is a measure of deformation from a harmonic confinement, and obtain \cite{golo06}
\begin{equation}
\frac{1}{2}{\bm h}(t)\cdot\mbox{\boldmath $\sigma$}=\frac{e\gamma\eta\hbar^2\omega_c}{9m^*\omega_0^4}
\left(E_y(t)\sigma_x+E_x(t)\sigma_y\right).
\label{eq12htseqEE}
\end{equation}
Finally, we note that the $p^3$-terms can also be relevant for spin relaxation
in quantum dots with anharmonic confining potential.
Of course, the magnetic field has to have an out-of-plane
component for this spin-electric coupling to dominate
over the one considered in the previous section.

\section{Spin--spin coupling via spin--orbit interaction}
\label{sec_5}

\begin{figure}[t]
\begin{center}
\includegraphics[angle=0,width=0.45\textwidth]{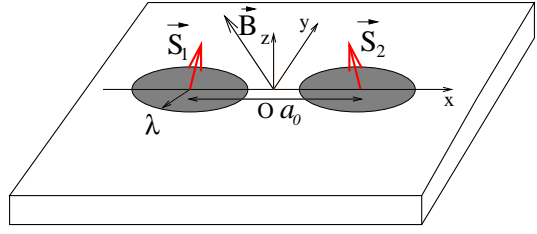}
\caption{(Color online) The figure shows a sketch  of the model  system which consists of two identical quantum dots in the $xy$-plane, separated by  distance $a_0$ (measured from dot-center to dot-center). $\vec{S}_i$ denotes the spin of electron $i=1,2$, $\lambda$ is the dot radius, and $\vec{B}$ is the external magnetic field.
The respective orbital wave functions of electron 1 and 2 are assumed to have no overlap
(i.e. tunneling between the dots is excluded). The remaining purely electrostatic
Coulomb interaction between the electron charges leads, via spin--orbit interaction, to
an effective coupling between their spins. This spin-spin interaction depends sensitively
on the  orientation  of $\vec{B}$, with no component along it, and is proportional to $\vec{B}^{2}$.}
\label{Quantumdots}
\end{center}
\end{figure}

In this part of the article, we discuss the interaction of two electron spins
localized in quantum dots through the combined effect of spin--orbit
interaction and Coulomb repulsion. The two single-electron quantum dot system
is shown in Fig.~\ref{Quantumdots}. It is assumed that the two dots are well
separated from each other such that there is no electron tunneling between
them. In this respect, the interaction between the spins is fundamentally
different from the Heisenberg exchange interaction for which the presence of
tunneling is crucial\cite{burk99}. Similarly, the combined effect of
Heisenberg exchange interaction and spin-orbit
coupling\cite{KVK,BSV,BuL,DBVBL,SB}is also based on tunneling and should be
carefully distinguished from the spin-orbit effect studied here. Even though the
Heisenberg exchange coupling allows typically for much stronger spin-spin
coupling than the electrostatically induced one\cite{trif2007}, the latter one
can prove useful for cases where it is difficult to get sufficient
wavefunction overlap (needed for large Heisenberg exchange), and, moreover, it
is also important to understand in detail the electrostatically 
induced spin--spin
coupling in order to get control over possible interference effects between
different types of coupling. This will be of importance for spin--qubit
applications in order to minimize spin decoherence and gate errors. 

We give now a short theoretical description of our system. The Hamiltonian of the two-single electron quantum dot system is
\begin{equation}
H=\sum_{i=1,2}\left(\frac{p_i^2}{2m^*}+U(r_i)+\frac{1}{2}g\mu_B\bm{B} \cdot \bm{\sigma}_i+H_{SO}^i\right)+\frac{e^2}{\kappa|\bm{r}_1-\bm{r}_2+\bm{a}_0|}\label{HamQDs} ,
\end{equation}
where the first two terms are the kinetic and orbital
confinement ($U(r_i)=m^*\omega_0^2r_i^2/2$), the third term is the Zeeman
energy, the fourth term stands for the spin--orbit interaction, both Rashba
and Dresselhaus [see Eq. (\ref{RDSOI})], while the last term stands for the
Coulomb coupling between the two electrons. The distance between the centers
of the two dots is $a_0$. Usually, the spin--orbit interaction is a weak
perturbation compared with the orbital level spacing and as a consequence can be
treated within perturbation theory, as it was done also in  the two
previous sections. However, here we have an additional energy scale given by
the strength of the Coulomb repulsion. This strength is measured through the
parameter $\delta=(\lambda/a_B)(\lambda/a_0)$\cite{trif2007}, where
$\lambda=\sqrt{\hbar/m^*\omega_0}$ is the dot radius and
$a_B=\hbar^2\kappa/m^*e^2$ is the Bohr radius in the material. In the general
case of arbitrary strong Coulomb repulsion, the effective spin Hamiltonian
$H_{spin}$ of the two electron system reads\cite{trif2007}         
\begin{equation}
H_{spin}=\frac{1}{2}E_{1Z}^{{\it eff}}\sigma_z^1+\frac{1}{2}E_{2Z}^{{\it eff}}\sigma_z^2+J_x\sigma_x^1\sigma_x^2+J_y\sigma_y^1\sigma_y^2,\label{effham}
\end{equation}
where the explicit expressions for the spin-orbit renormalized  Zeeman
splittings $E_{iZ}^{{\it eff}}$ and spin-spin couplings $J_{x,y}$ are given in
Ref.~\citeonline{trif2007}. 
This interaction vanishes for vanishing  Zeeman
splitting  and is highly anisotropic (XY type). Due to the finite Zeeman
splitting, the relevant electrostatically induced spin-spin coupling can be
written as  
\begin{equation}
H_{s-s}=J_{{\it eff}}(\sigma_+^1\sigma_-^2+\sigma_+^2\sigma_-^1),\label{spineffective}
\end{equation}
with $\sigma_{\pm}=\sigma_x\pm i\sigma_y$ and $J_{{\it
    eff}}=(1/2)(J_x+J_y)$\cite{trif2007}. Up to now we posed no assumptions on
the strength of the Coulomb repulsion. However, there are two interesting
limiting cases, namely $\delta\ll1$ (weak Coulomb
repulsion) and $\delta\gg1$ (strong Coulomb repulsion). 
 
\begin{figure}[t]
\begin{center}
\includegraphics[angle=0,width=0.45\textwidth]{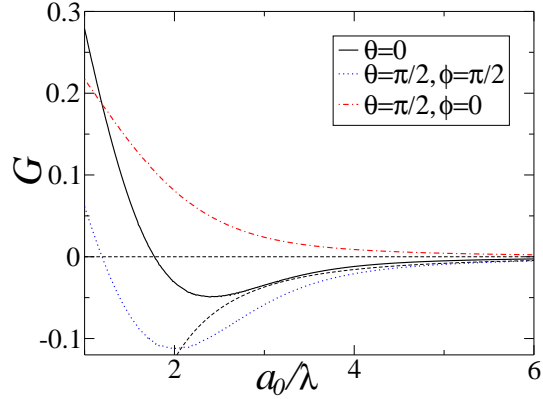}
\caption{(Color online) The function $G$ occurring  in Eq. ($\ref{couplingpert}$) plotted  as a function of the geometric 
distance $a_0$ between the dot centers scaled by the dot radius $\lambda$ for different magnetic field orientations. 
The dashed line represents the dipolar approximation of $G$ for a perpendicular magnetic field 
($\theta=0$) which scales like $a_0^{-3}$.}
\label{Gfunction}
\end{center}
\end{figure}

In the first case, $\delta\ll1$, the Coulomb interaction is a weak perturbation compared to the bare orbital level spacing $\hbar\omega_0$ such that 
\begin{equation}
H_{s-s}=\int d\bm{r}_1d\bm{r}_2\frac{\delta\rho_1\delta\rho_2}{\kappa|\bm{r}_1-\bm{r}_2+\bm{a}_0|}.\label{Coulombpert}
\end{equation}
Here, the 2x2 matrices $\delta\rho_{1,2}$ are the  spin-orbit induced charge
distributions or 
spin-dependent charge distributions in each dot in the absence of Coulomb
interaction\cite{trif2007}. From Eq. (\ref{Coulombpert}) we see that the
spin-spin interaction results from a Coulomb-type coupling between two
charge distributions which themselves depend on spin. In the limit of large
interdot distances $a_0\gg\lambda$, we can perform a multipolar expansion,
such that within the lowest order we obtain 
\begin{equation}
H_{s-s}\approx\frac{\bm{m}_1\cdot\bm{m}_2-3(\bm{m}_1\cdot\bm{n}_a)(\bm{m}_2\cdot\bm{n}_a)}{\kappa a_0^3}\label{dipolar}
\end{equation}     
where $\bm{n}_a=\bm{a}_0/a_0$. The dipole moments
$\bm{m}_i=\langle0|\delta\rho_i\bm{r}_i|0\rangle\equiv\bar{\bar{\mu}}\bm{\sigma}_i$,
where $|0\rangle$ is the orbital ground-state and $\bar{\bar{\mu}}$ is the
tensor corresponding to an effective spin-orbit magneton (for explicit
expressions see Ref.~\citeonline{trif2007}). The strength of this effective
spin-orbit induced magneton is given by $||\bar{\bar{\mu}}||\approx
eE_Z/m^*\omega_0^2\lambda_{SO}$.  To give an estimate, we assume
$\hbar\omega_0\sim 0.5\,{\rm meV}$, $E_Z\sim 0.05\, {\rm meV}$
($B\sim 2\, {\rm T}$)  and $m^*=0.067 m_e$, $\lambda_{SO}\sim
10^{-6}\,{\rm m}$ for GaAs quantum dots which gives, when compared with the
Bohr magneton, $||\bar{\bar{\mu}}||/\mu_B\sim 10^3$. This implies that the
spin-orbit induced dipole-dipole interaction in Eq. (\ref{dipolar}) can be
much stronger than the direct dipole-dipole interaction in vacuum, whose
strength is given by $\mu_B$. Also, still in the limit $\delta\ll1$, but for
arbitrary interdot distance $a_0$ the effective coupling $J_{{\it eff}}$ has
the form 
\begin{equation}
J_{{\it
    eff}}=E_Z\frac{\lambda}{a_B}\frac{E_Z}{\hbar\omega_0}\left(\frac{\lambda}{\lambda_{SO}}\right)^2G(a_0/\lambda,\theta,\Phi)\label{couplingpert} ,
\end{equation}        
where the function $G(a_0/\lambda,\theta,\Phi)$  is plotted in
Fig.~\ref{Gfunction} as a function of $a_0/\lambda$ for different angles
$\theta,\Phi$. The key feature of the electrostatic  spin-spin interaction is
that it can range from ferromagnetic to antiferromagnetic type, depending on
the magnetic field orientation, passing even through zero for certain angles
and/or inter-dot distances.    

We now focus on the opposite limit $\delta\gg1$, when the Coulomb interaction
is much stronger  than the bare orbital level spacing $\hbar\omega_0$. Then,
we approximate $e^2/\kappa|\bm{r}+\bm{a}_0|\rightarrow (e^2/2\kappa
a^3)[3(\bm{n}_a\cdot\bm{r})^2-r^2]$\cite{trif2007}, where 
$a$ is the effective distance between the electrons  due to the combined
effect of Coulomb repulsion and orbital confinement. For the explicit
derivation of the effective distance $a$ in terms of the bare one $a_0$ see
Ref.~\citeonline{trif2007}. Within this ansatz, the spin-spin coupling
Hamiltonian takes the form 
\begin{equation}
H_s=\frac{E_Z^2}{m^{*2}\omega_0^2\lambda_{SO}^2}\left[\left(\frac{1}{b_x^2}-1\right)\sigma_x^1\sigma_x^2+\left(\frac{1}{b_y^2}-1\right)\sigma_y^1\sigma_y^2\right]
\label{62}
\end{equation}
for the case of a perpendicular magnetic field. 
In the above expression we have $b_x=\sqrt{1+4(\lambda/a_B)(\lambda/a)^3}$ and
$b_y=\sqrt{1-2(\lambda/a_B)(\lambda/a)^3}$.  

Let us give now some estimates for the coupling $J_{{\it eff}}$ when an in-plane magnetic field is applied along, say,
the $x$-direction. 
Assuming now GaAs quantum dots, and  $E_Z=0.1\; {\rm meV}$ ($B=4 {\rm \; T}$), $\hbar\omega_0=0.5\; {\rm meV}$ 
($\lambda/a_B\sim 5$), $\lambda/\lambda_{SO}\sim 10^{-1}$. 
Using these numbers and taking for the geometric inter-dot distance
$a_0/\lambda\sim 2$, we obtain 
$J_{{\it eff}}\sim 10^{-7}\; {\rm eV}$. 
It is worth mentioning that the hyperfine interaction between the electron and the collection of nuclei in a quantum dot 
($\sim 10^5$) leads to similar energy scales\cite{khae02,cois04}. 
This shows that the spin-spin coupling derived here can be very relevant for the spin dynamics in the case of 
electrostatically coupled quantum dots and that it can also compete with other types of interactions. 
Considering now the case of InAs quantum dots \cite{fasth05,pfund07} 
in a magnetic field along the $x$ direction, with 
$\lambda_{SO}\sim 2\lambda\sim 100 {\rm nm}$ and $E_Z/\hbar\omega_0=0.1$ and
taking also $a_0/\lambda\sim 2$, 
a value of $J_{{\it eff}}\sim 10^{-6}{\rm eV}$ is obtained. 

\section{Conclusions}
\label{sec_6}

We have discussed several selected topics on the theory of spin qubits in
nanostructures. We have first reviewed our recent proposal how to form spin
qubits in graphene. This is interesting for two reasons. On the one hand, one
expects very long spin lifetimes in graphene because of a weak spin--orbit
interaction and very few host atoms with a nuclear spin. On the other hand,
spin qubits in graphene allow for a new type of long distance coupling that
uses the property that a ribbon of graphene is a small bandgap
semiconductor. Furthermore, we have pointed out several aspects of spin
relaxation and decoherence due to spin--orbit interaction and the coupling to
a bath. As two possible dissipation channels we have considered lattice
vibrations (phonons) and charge fluctuations in the surrounding environment,
for instance, a nearby quantum point contact. Subsequently, we have shown how
to use EDSR to rotate the spin of an electron in a
quantum dot using an oscillating electric field (instead of the oscillating
magnetic field employed in the usual ESR). In the final
part of the review article, we have discussed how to couple two spins (located
in two different quantum dots) via
spin--orbit interaction in a situation in which direct 
tunneling between the dots is
highly suppressed. 

We would like to thank D.V. Bulaev, G. Burkard, and V.N. Golovach for the
collaboration on the work reviewed in this article. Financial support has been
provided by the Swiss NSF, the NCCR Nanoscience, and JST ICORP.


\begin{thebibliography}{99}
\bibitem{foot1}
In Eq.~(\ref{bc}), the semiconducting armchair boundary conditions
are formulated in a slightly different way than in
Ref.~\citeonline{brey06} which is due to a different definition of the
width $W$ of a ribbon of graphene. 
In Ref.~\citeonline{brey06}, the width $W$ is defined as the distance from
one outer edge of the armchair boundary to the other outer edge of the
armchair boundary (on the other side of the ribbon). In Ref.~\citeonline{trau07}
(and also in Ref.~\citeonline{twor06}) instead, the width is defined as the distance
from one inner edge of the armchair boundary to the other inner edge of the
armchair boundary (on the other side of the ribbon). These boundary conditions
are fully consistent with each other and with an earlier tight-binding model
analysis of the spectrum of graphene nanoribbons. \cite{naka96}
\bibitem{foot2}
An alternative way to lift the valley degeneracy in graphene (which does not
rely on special boundary conditions) is to form quantum dots in ring structures
and to apply a magnetic field perpendicular to the surface of graphene, see
Ref.~\citeonline{recher}.
\end{thebibliography}
\end{document}